\documentclass{IEEEoj}
\usepackage{cite}
\usepackage{amsmath,amssymb,amsfonts}
\usepackage{algorithmic}
\usepackage{graphicx,color}
\usepackage{textcomp}
\usepackage{subcaption}
\usepackage{comment}
\usepackage{listings}
\usepackage{makecell}
\usepackage[normalem]{ulem}
\usepackage{hyperref}

\def\BibTeX{{\rm B\kern-.05em{\sc i\kern-.025em b}\kern-.08em
    T\kern-.1667em\lower.7ex\hbox{E}\kern-.125emX}}
\makeatletter
\def\ps@headings{\def\@oddhead{\vbox{\vspace{17pt}\hsize\textwidth\hbox{\rfxfont\rightmark\hfill}\hfill\par
\smallskip\noindent\hbox to \textwidth{\vrule width\textwidth height.3pt depth0pt}}}%
\def\@evenhead{\vbox{\vspace{17pt}\hsize\textwidth\hfill\hbox{\hfill\rhfont\leftmark}\par
\smallskip\noindent\hbox to \textwidth{\vrule width\textwidth height.3pt depth0pt}}}%
\def\@oddfoot{\hfill\rffont\thepage}\def\@evenfoot{\rffont\thepage\hfill}}
\def\ps@plain{\def\@oddhead{\vbox{\vspace{17pt}\hsize\textwidth\hbox{\rhfont\leftmark\hfill}\hfill\par
\smallskip\noindent\hbox to \textwidth{\vrule width\textwidth height.3pt depth0pt}}}%
\def\@evenhead{\vbox{\vspace{17pt}\hsize\textwidth\hfill\hbox{\hfill\rhfont\leftmark}\par
\smallskip\noindent\hbox to \textwidth{\vrule width\textwidth height.3pt depth0pt}}}%
\def\@oddfoot{\hfill\rffont\thepage}\def\@evenfoot{\rffont\thepage\hfill}}
\makeatother
\AtBeginDocument{\definecolor{ojcolor}{cmyk}{1,0.45,0,0.18}%
\definecolor{ojcolor2}{cmyk}{0,0.91,0.81,0.19}%
\pagestyle{headings}\thispagestyle{plain}}
\begin{document}

\title{\textcolor{ojcolor2}{SpecPCM: A Low-power PCM-based In-Memory Computing Accelerator for Full-stack Mass Spectrometry Analysis}}

\author{KEMING FAN\authorrefmark{1},
ASHKAN MORADIFIROUZABADI\authorrefmark{1}, 
XIANGJIN WU\authorrefmark{2}, 
ZHEYU LI\authorrefmark{1}, 
FLAVIO PONZINA\authorrefmark{1}, 
ANTON PERSSON\authorrefmark{2}, 
ERIC POP\authorrefmark{2}, 
TAJANA ROSING\authorrefmark{1}, 
MINGU KANG\authorrefmark{1}}
\affil{\authorrefmark{1}University of California San Diego, San Diego, CA 92093 USA} \affil{\authorrefmark{2}Stanford University,  CA 94305 USA}
\corresp{CORRESPONDING AUTHOR: M. KANG (e-mail: mingu@ucsd.edu).}
\authornote{This work was supported by the PRISM Center within JUMP 2.0, a Semiconductor Research Corporation (SRC) program  with DARPA.
\vspace{-1cm}}

\IEEEpubid{0000--0000/00\$00.00~\copyright~2021 IEEE}

\begin{abstract}
Mass spectrometry (MS) is essential for proteomics and metabolomics but faces impending challenges in efficiently processing the vast volumes of data. This paper introduces SpecPCM, an in-memory computing (IMC) accelerator designed to achieve substantial improvements in energy and delay efficiency for both MS spectral clustering and database (DB) search. SpecPCM employs analog processing with low-voltage swing and utilizes recently introduced phase change memory (PCM) devices based on superlattice materials, optimized for low-voltage and low-power programming.  Our approach integrates contributions across multiple levels: application, algorithm, circuit, device, and instruction sets. We leverage a robust hyperdimensional computing (HD) algorithm with a novel dimension-packing method and develop specialized hardware for the end-to-end MS pipeline to overcome the non-ideal behavior of  PCM devices. We further optimize multi-level PCM devices for different tasks by using different materials.
We also perform a comprehensive design exploration to improve energy and delay efficiency while maintaining accuracy, exploring various combinations of hardware and software parameters controlled by the instruction set architecture (ISA). SpecPCM, with up to three bits per cell, achieves speedups of up to 82× and 143× for MS clustering and DB search tasks, respectively, along with a four-orders-of-magnitude improvement in energy efficiency compared with state-of-the-art CPU/GPU tools.

\end{abstract}


\maketitle

\section{Introduction}
\IEEEPARstart{M}{ass} spectrometry (MS) is a key analytical tool used by proteomics and metabolomics, aiding in drug discovery and chemical analysis by identifying and quantifying molecules based on their mass-to-charge ratios\cite{ wilhelm2014mass}. Its high sensitivity and precision have made it one of the most widely used techniques for detecting even the smallest molecular variations.
However, one of the key challenges in MS processing is handling the vast and continually growing data volumes. For example, the MassIVE database, a publicly accessible repository for proteomics MS data \cite{MassIVE}, now exceeds 600 TB (as of September 2024) and continues to grow at an accelerating pace, with hundreds of terabytes of new spectral data added annually.
The MS analysis process involves comparing spectra generated from MS experiments against an extensive reference library to identify proteins, a procedure known as database (DB) search~\cite{lam2011building}. To accelerate the search process, spectral clustering is done by grouping similar reference spectra together. During search, the query is first compared to the cluster centroids, quickly focusing the search on an appropriate cluster and thereby speeding up the overall process~\cite{griss2016recognizing}.
Ideally, such a database should be clustered on a daily basis as new samples are continually added, but this is currently done only once per year due to the excessive time required, resulting in lower accuracy.
Traditional systems with separate memory and processor units are hindered by limited data movement bandwidth and computing efficiency in processing such a sheer amount of data. Modern MS analysis tools \cite{ANN_SoLo'23, xu2023hyperspec, kang23hyperoms}, whether based on conventional CPU or GPU architectures, often spend more than 60\% of their time on large matrix operations with significant memory footprint. These tools struggle to manage large datasets efficiently, mainly due to the high energy consumption and latency involved in data transfer.

To overcome these challenges, in-memory computing (IMC) has emerged as an alternative paradigm that processes data directly within the memory where it is stored, substantially reducing the latency and energy overhead caused by data movement.
To capitalize on this opportunity, various memory topologies have been explored, including DRAM, SRAM, RRAM, PCM, NAND Flash, and other emerging memory technologies~\cite{shanbhag2022comprehending}.
While RRAM has been widely adopted \cite{wan2022compute, fan2024efficient, wan2020voltage} for its well-recognized high-density and efficient read operations, particularly due to its support for multi-level cells (MLC), it faces limitations, such as high energy consumption and high voltage requirements during write operations. This drawback will significantly degrade the energy efficiency of the clustering process, where frequent data updates are required to adapt to the newly collected MS data.
While NAND flash memories provide superior memory density and fabrication maturity, they suffer from relatively high latency, e.g., several $\mu s$\cite{hsu2023storage}, due to the high resistance in the read-path, which is caused by the nature of reading data from serially connected cells.

This work adopts a recently developed multi-level phase change memory (PCM)~\cite{wu2024pcm} based on superlattice materials, which features lower error rates, reduced voltage requirements,  faster and more energy-efficient programming.
In particular, we aim to explore the analog IMC, which offers dramatic efficiency improvements, achieving more than two orders of magnitude benefit~\cite{shanbhag2022comprehending} compared to conventional digital counterparts by utilizing low-voltage swing analog operations. Additionally, the analog IMC performs both reading and computation across the entire bitcell array simultaneously, enabling high parallelism and significant compute density improvement. Consequently, the analog IMC on PCM facilitates efficient processing for classification while also enabling the effective updating of stored weights to adapt to newly collected MS data.

From an algorithmic perspective, we employ hyperdimensional computing (HD), a brain-inspired computing paradigm that leverages lightweight and highly parallel operations by encoding input features into high-dimensional (long) binary vectors. HD replaces costly and complex floating-point arithmetic with simpler binary or integer operations, which can be executed in parallel, leading to dramatic throughput improvements as demonstrated in \cite{karunaratne2020memory, kang2022openhd, xu2023fsl}. 
Furthermore, HD’s data representation in hyperspace offers significant error resilience, with data points being well separated by large geometric distances. This property has been demonstrated in previous work \cite{fan2024efficient}, where HD tolerated up to a 10\% bit error rate for MS DB search tasks.  This resilience creates a strong synergy with analog IMC on emerging devices, helping to overcome their computing and storage non-idealities while achieving greater storage density and computing efficiency.

The use of MLC PCM involves complex trade-offs between efficiency and accuracy, which require careful optimization across both hardware and algorithms. These interrelated challenges cannot be addressed at a single abstraction level. Therefore, we introduce SpecPCM, an IMC accelerator designed for the efficient processing of MS workloads. This framework integrates design efforts across the entire vertical stack, spanning application, algorithm, circuit, device, and instruction set levels, to enhance performance throughout the end-to-end MS pipeline.
The detailed  contributions of this work are summarized as follows:
\begin{enumerate}
\item{We propose an analog IMC system with architecture and circuits specifically tailored for MS algorithms. While prior works have applied IMC to MS database (DB) search tasks in the HD domain, our work is the first to apply IMC for both clustering and DB search.}
\item {At the algorithm level, we introduce a new HD encoding method, called \textit{dimension packing}, to maximize storage density by leveraging multi-level PCM devices while maintaining the simplicity of the binary representation of HD vectors.}
\item {At the device level, we propose customized PCM devices to meet the distinct requirements of clustering and MS DB search by optimizing the materials differently for each task, based on measured characterization results from the fabricated devices.}
\item {We conduct hardware-software co-design through a comprehensive analysis to balance trade-offs between latency, energy efficiency, and accuracy, taking into account various parameters such as bits per cell, write-verify cycles, analog-to-digital converter (ADC) precision, and HD dimensions, all controlled by the instruction set.}

\end{enumerate}

The results indicate that the proposed SpecPCM demonstrates speedup of up to 82$\times$ for clustering and 143$\times$ for database search over state-of-the-art (SoA) solutions, with a four-orders-of-magnitude in energy efficiency improvement, while maintaining on-par accuracy across datasets of different scales.

\section{Background and Motivation}

This section provides background on HD, the MS algorithm, and the IMC for the MS analysis to motivate the proposed PCM-based analog IMC architecture.

\subsection{Hyperdimensional computing}
Hyperdimensional computing (HD) is a brain-inspired computing paradigm that leverages lightweight and highly parallel operations, by encoding input features into high-dimensional binary vectors called hypervectors (HV), typically with a dimension of 1k-10k ~\cite{karunaratne2020memory}.

\noindent \textbf{HD Encoding:} Encoding is the process of mapping raw data to an HD vector. Despite its diversity,  we introduce the ID-level encoding~\cite{fan2024efficient} approach.
Initially, two sets of hypervectors (HVs) are created, ID HVs and level HVs. Both sets consist of $D$-dimensional HVs, where each element is either -1 or 1. The encoding scheme assigns a unique ID HV to each feature position. These ID HVs are randomly generated to ensure orthogonality among all features. Additionally, a set of level HVs is generated to represent the value of each feature. 
To create these level HVs, we identify the minimum and maximum feature values across all data points, denoted as $l_{min}$ and $l_{max}$. The range $[l_{min}, l_{max}]$ is then quantized into $m$ levels. Each level is associated with a corresponding level HV, e.g.,  $LV_1$, $\ldots$, $LV_m$ for the $m$ different levels.
To encode a feature vector, the encoder performs an element-wise multiplication and accumulation (MAC) between the position ID HV ($ID_i$) and the corresponding level HV ($hv_i$) for the $i$-th feature position. The resulting hypervector ($HV_i$) is then binarized to complete the encoding process, i.e., $HV_i = sign(ID_i \cdot hv_i)$. The following equation demonstrates how a $D$-dim feature vector is mapped into the HD space.
\begin{equation}
    HV_i 
=  sign\left(hv_{i1}*ID_{i1} + \cdots + hv_{iD}* ID_{iD} \right) 
\label{eq: encoding}
\end{equation}
where $hv_i \in \{LV_1,  \ldots, LV_m\}$, $LV_k \in \{-1, 1\}^D$, $ID_i \in \{-1, 1\}^D$, and $sign$ 
outputs 1 when the input is positive and -1 otherwise. 
The $ID_{ij}$ and $hv_{ij}$ are the $j$-th element of $ID_i$ and $hv_i$, respectively.

%

\noindent 
\textbf{Classification and Inference:}
The HD model classifies input samples using class HVs, each representing a specific class. Classification involves calculating the similarity between the encoded query and class HVs, with the predicted class being the one whose HV is most similar to the input.
Similarity is computed using Hamming distance, which equals the dot product of two bipolar vectors, and the class with the highest score is chosen:
$y = argmax_j \big( \sigma(\vec{\mathcal{Q}}, \vec{\mathcal{C}_j})\big)$, where $ \sigma$ represents the similarity function, $\vec{\mathcal{Q}}$ is the encoded query vector, $\vec{\mathcal{C}}$ is the class HV, and $y$ is the predicted label.

\subsection{Mass Spectrometry and its accelerations}

Proteomics is essential for understanding biological processes and drug discovery, as it identifies key proteins, biomarkers, and therapeutic targets involved in health and disease. Mass spectrometry (MS) is a key tool in this field, enabling detailed analysis of protein compositions. MS measures the mass-to-charge ratio (m/z) of ionized proteins, with the results presented as a spectrum—a plot of intensity as a function of the m/z ratio.
Modern MS experiments generate millions of spectra, requiring the efficient processing of hundreds of terabytes (TB) of data \cite{MassIVE}.
MS analysis involves two main tasks: clustering and database (DB) search, both of which can be implemented using HD.
In the clustering process (Fig.~\ref{clus_intro}), spectra are first divided into several buckets based on bio-features.
Inside each bucket, spectra are encoded into HVs, and pairwise distances between data points are calculated using dot products to form a distance matrix. The algorithm begins with each data point in its own cluster and iteratively merges the closest clusters until a distance threshold is reached. This process generates the representative reference database with a condensed volume of data.
For the DB search (Fig.~\ref{DBsearch_intro}), a Hamming similarity search compares query HVs to reference HVs in the DB, identifying the closest reference based on the dot product score. Finally, matching candidates are filtered with a false discovery rate (FDR), a common technique using decoy spectra to evaluate accuracy~\cite{elias2007target}.

In both clustering and DB search, each spectrum from an MS experiment is extensively compared against a collection of spectra, a process that is particularly time-consuming due to the sheer volume of data involved.
Previous works, such as \cite{bittremieux2021falcon}, \cite{wang2018mscrush}, and ANN-SoLo \cite{ANN_SoLo'23}, have attempted to accelerate MS analysis using techniques including hashing, approximate nearest neighbor search, and efficient dot products. However, these tools have limited efficiency due to their reliance on complex, high-precision floating-point arithmetic.
In contrast, HD-powered tools, such as HyperSpec \cite{xu2023hyperspec} for clustering and HyperOMS \cite{kang23hyperoms} for DB search, demonstrate the fastest operations by utilizing simple Boolean operations and enabling significantly higher hardware parallelism.

\begingroup

\begin{figure}
\centerline{\includegraphics[width=0.46\textwidth]{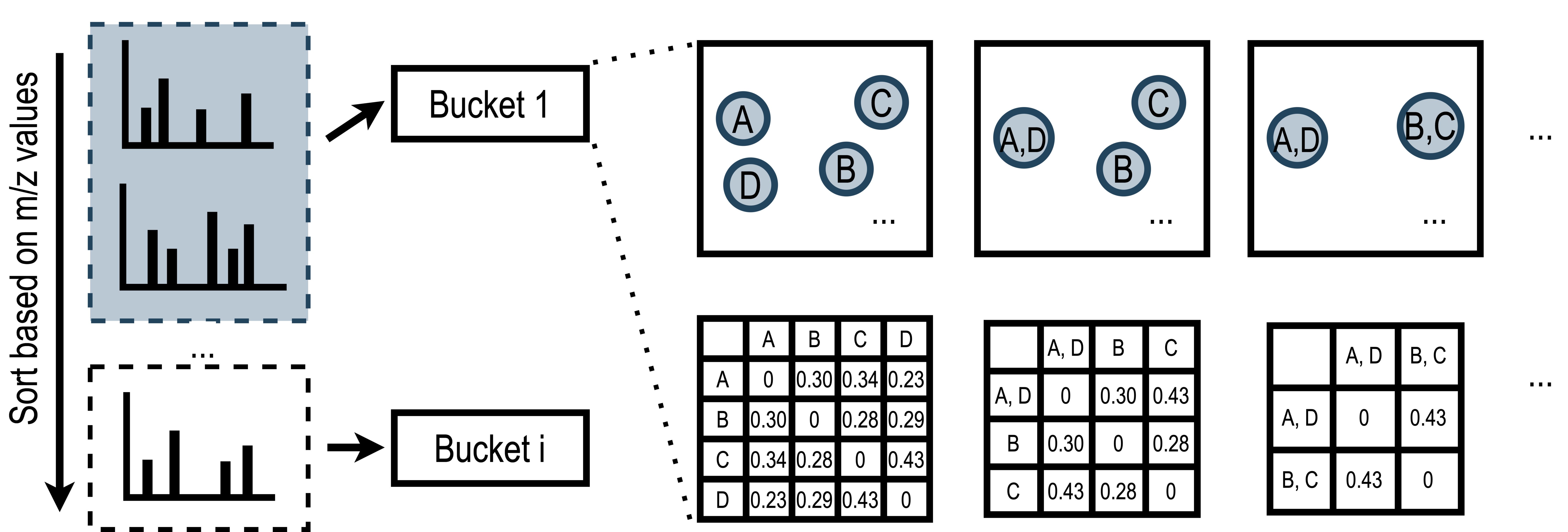}}
\caption{Overview of spectral clustering for MS analysis.}
\label{clus_intro}
\vspace{-0.6cm}
\end{figure}

\begin{figure}
\centerline{\includegraphics[width=0.48\textwidth]{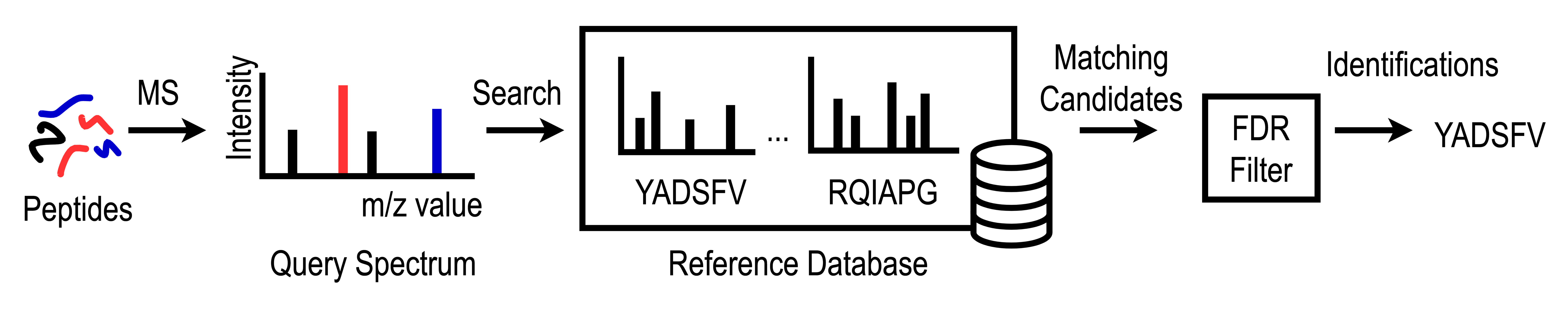}}
\caption{Overview of MS database (DB) search.}
\label{DBsearch_intro}
\vspace{-0.6cm}
\end{figure}

Despite the latency improvements provided by HD, latency profiling results (Fig.~\ref{motivation}) reveal a new scalability challenge when handling large datasets. Even with an NVIDIA 4090 GPU equipped with 24 GB VRAM, distance calculation remains the primary bottleneck in clustering, while Hamming similarity search is the main bottleneck in DB search. Both stages involve large-scale matrix computations, leading to significant data movement, particularly when the dataset exceeds the GPU's onboard memory capacity. This data movement between the GPU and main memory hampers processing efficiency and results in substantial performance overheads.
These observations underscore the motivation for accelerating key operations, such as distance calculation and Hamming similarity search, using analog IMC. By processing directly within memory in combination with HD, this approach aims to minimize the costly data movement and enhance the overall efficiency of MS analysis. %

\begin{figure}[h]
\centerline{\includegraphics[width=3.5in]{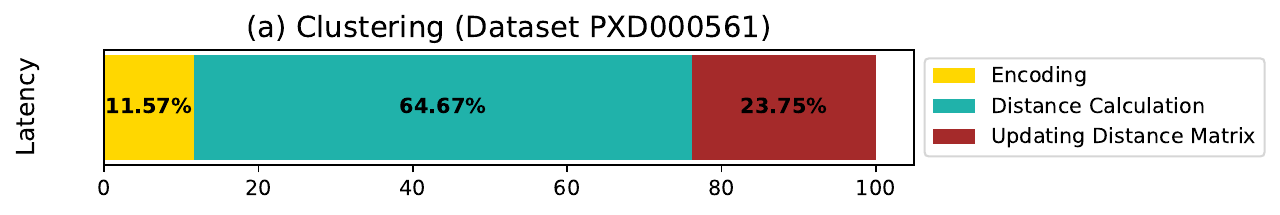}}
\centerline{\includegraphics[width=3.5in]{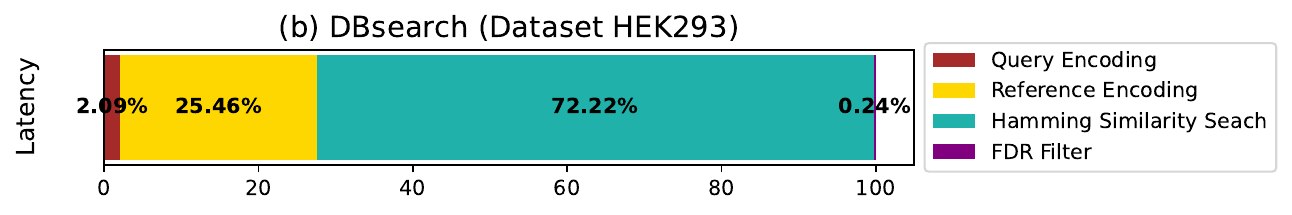}}
\caption{Latency breakdown for GPU tools, (a) Clustering on dataset PXD000561 using HyperSpec, (b) Database search on dataset HEK293 using HyperOMS.}
\label{motivation}
\vspace{-1.2cm}
\end{figure}

\subsection{IMC for MS analysis and application of PCM}

The IMC has been exploited to accelerate MS analysis using various memory technologies. While DRAM~\cite{kang2024dram} and NAND flash~\cite{hsu2023storage} have been employed in near-memory computing architectures for MS analysis, they are rarely used for analog IMC. This is due to DRAM's volatility and NAND flash's serially-connected structures, which prevent the analog IMC operations from all bitcells.
In contrast, RRAM has been actively explored for analog IMC in MS analysis in conjunction with HD, due to its fast read access, simple fabrication process, and suitability for multi-level cell operations, primarily focusing on classification for DB search \cite{fan2024efficient}. Despite these advantages, RRAM suffers from orders of magnitude higher write latency compared to read latency and the requirement for significantly high programming voltages (e.g., more than 2 V). 
These drawbacks are less apparent during classification tasks, where the weights remain static after the model is trained. However, in tasks such as training and spectral clustering, frequent data updates are required, necessitating repeated write operations, which exacerbate the impact of these limitations.
Unlike previous studies that focus either on clustering or DB search, this work aims to provide an end-to-end solution for both. To achieve this, we adopt a recently introduced PCM device with a superlattice structure based on nanocomposites of Ge$_4$Sb$_6$Te$_7$ \cite{wu2024pcm}  to leverage the following unique advantages: 1) a very low programming voltage (less than 1.0 V), which ensures great compatibility with a logic die without requiring specialized peripheral circuitry to support high voltage; 2) low switching energy (approximately pJ) due to the low forming voltage; 3) reduced resistance drift, which significantly supports stable multi-level cells (MLCs) and increases the effective storage density; and 4) PCM's well-established technology and process maturity.

Despite the above benefits, utilizing MLC on PCM remains challenging due to increased susceptibility to noise and variability, with error rates often exceeding 10\% even after meticulous write-verify operations~\cite{fan2024efficient}. These challenges impede real-world deployment, underscoring the need for error-tolerant algorithms. This work leverages HD computing for its superior error resilience, as demonstrated in  \cite{fan2024efficient}.

\section{Proposed SpecPCM using Analog IMC based on PCM}\label{sec: proposed}
This section outlines the hardware and software implementations of SpecPCM, covering algorithmic, architectural, and device perspectives.

\subsection{SpecPCM overview} 
\begin{figure}
\centerline{\includegraphics[width=3.5in]{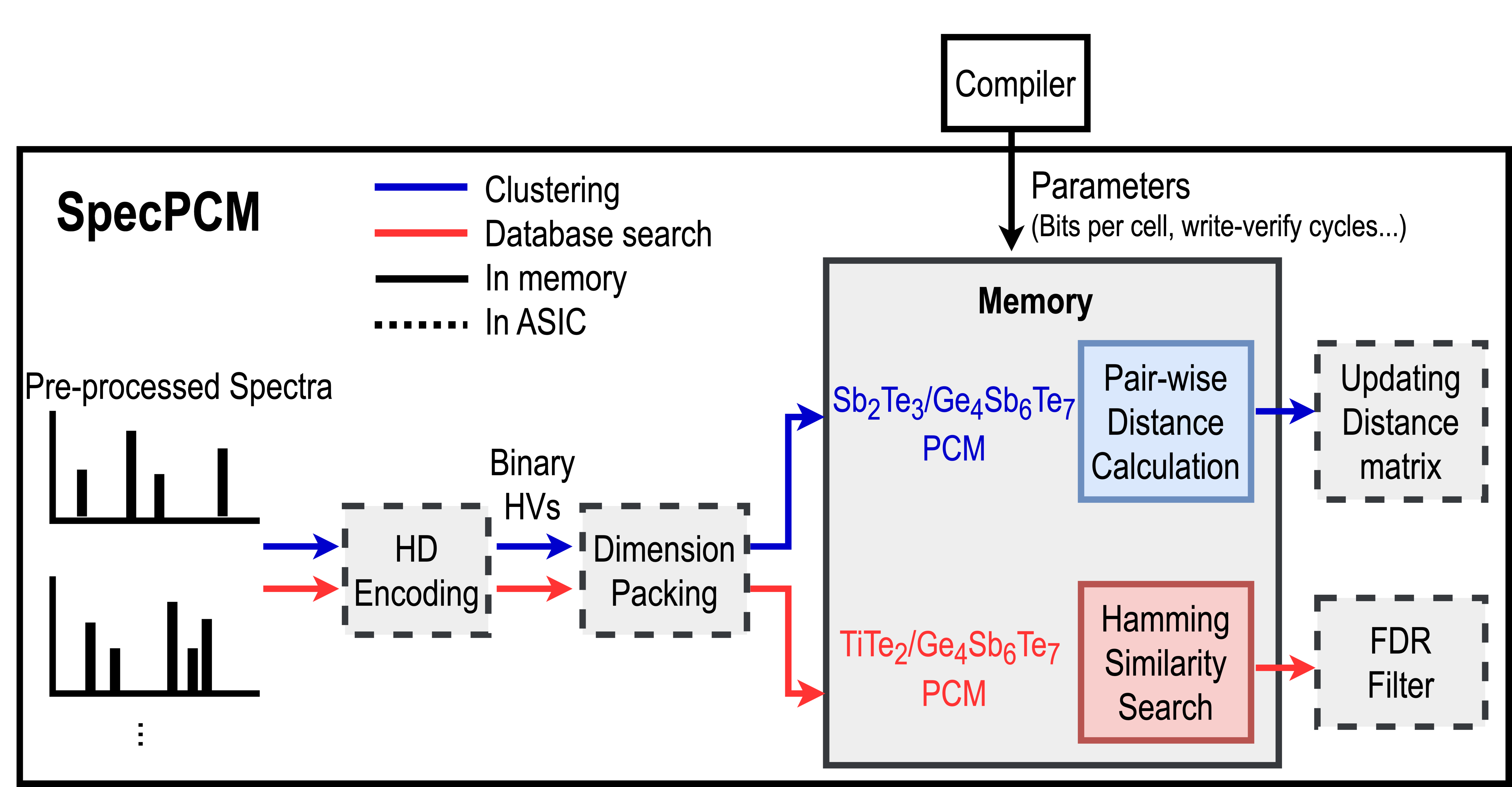}}
\caption{Overview of the SpecPCM accelerator.}
\label{overview}
\vspace{-0.3cm}
\end{figure}

\begin{figure}
\centerline{\includegraphics[width=0.4\textwidth]{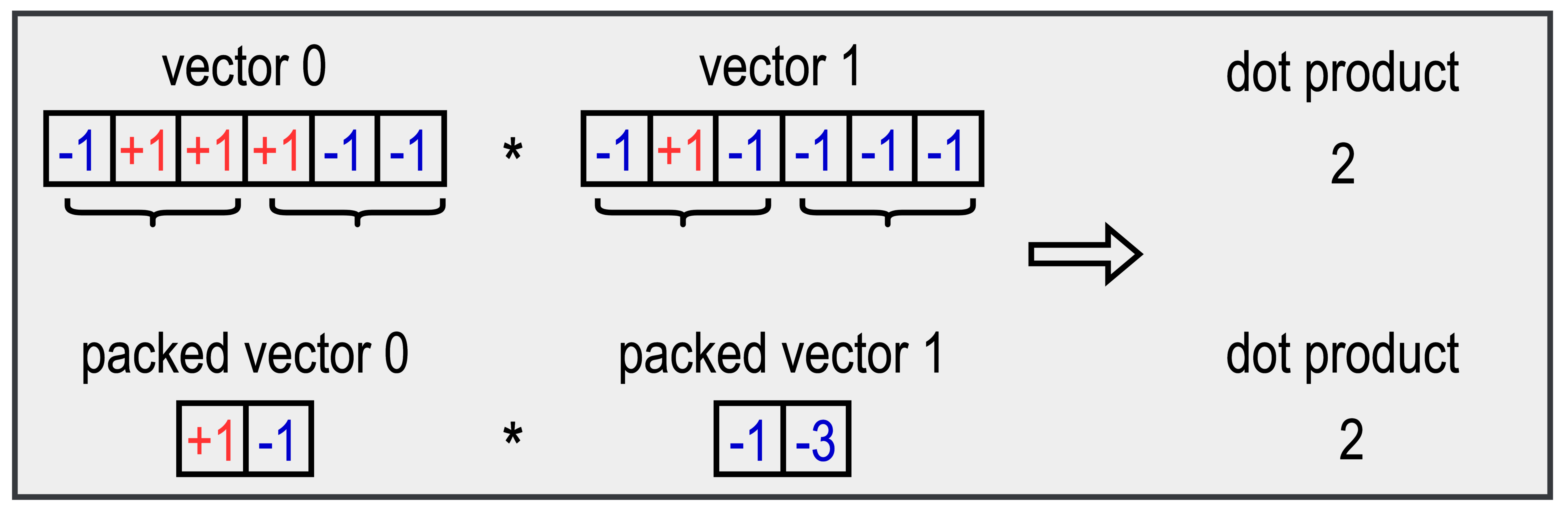}}
\caption{Dimension packing for MLC with 3 bits per cell.}
\label{fig:packing}
\vspace{-0.5cm}
\end{figure}

Figure \ref{overview} illustrates the system overview of the SpecPCM accelerator, which supports both clustering and DB search. The memory-intensive distance calculations in clustering and Hamming distance calculation in DB search,  identified as the main bottlenecks, are offloaded to PCM-based memory for the IMC processing. The remaining steps are handled by near-memory ASIC blocks.
Encoded HVs are packed and then passed to memory. Each memory bank array is sized as 128$\times$128 cells and 2T2R per cell, with multiple arrays to enable parallel processing. 
For clustering, distance calculations are performed directly in Sb$_2$Te$_3$/Ge$_4$Sb$_6$Te$_7$ PCM, which supports lower programming power whereas hamming similarity searches are executed in TiTe$_2$/Ge$_4$Sb$_6$Te$_7$ PCM, which offers longer retention time and lower error rates ~\cite{wu2024pcm}. 
A detailed discussion of the selection of these PCM types is provided in Section~\ref{sec: pcm devices}.
Various parameters for controlling IMC operations, including the number of bits per cell, write-verify cycles, and ADC precision, are managed by the software through a custom instruction set architecture (ISA), which will be discussed in Section~\ref{sec: compiler}. 

\subsection{Dimension packing in HD algorithm}

Previous work has shown that ID-level encoding performs best for MS analysis \cite{kang23hyperoms, xu2023hyperspec} as it captures essential spectral information with minimal loss. 
Each spectrum is encoded as a HV in binary format, as shown in Equation (\ref{eq: encoding})
and later operations (e.g., dot product) on those binary vectors require bit-wise operations. In contrast, MLC hardware is designed for non-binary, integer-based computations. Therefore, directly storing and processing binary vectors on MLC devices is suboptimal in terms of storage and computational efficiency.
To address this, we propose a dimension packing method, which is processed in ASIC. This involves converting the binary vector of length $D$ into a compressed vector of length $D/n$ by summing $n$ adjacent bits, where $n$ is the number of bits per cell. This compression aligns the data with the MLC devices' optimal format, improving the storage and compute density by $n$ times while maintaining accuracy with only a negligible drop compared to the original binary version, as will be shown in Section~\ref{sec: evaluation}.

\subsection{Hardware Operations}\label{sec: hardware}
\begingroup
The clustering begins with the encoding and dimension packing of spectral data. 
These packed HVs are then programmed into the PCM memory arrays, each configured as 
a 128×128 matrix using a 2T2R cell structure for each element.
As shown in Fig.~\ref{fig: cim_array}, each element is represented by a pair of PCM cells to store signed numbers, with the value expressed as the difference in conductance between the two cells, as introduced in \cite{wan2022compute, wan2020voltage}.
Each row in the array contains the data for a single HV, with multiple HVs stored across different rows in the array.
Due to the high dimensionality of the HVs (e.g., 1024 dimensions after packing), a single row in the array cannot store an entire HV. Instead, each row in an array stores a different segment of HV, with parts of the same HV distributed across multiple arrays at the same row. Multiple arrays can operate in parallel for higher throughput.

\noindent \textbf{Programming:} To program the HVs into the array, pulses  are generated by the generator module and transmitted to the array through source-line (SL) drivers. 
The voltage level of each SL is modulated based on the target resistance value of each cell inside the write pulse generator.
The bit-lines (BLs) are connected to the ground and the target row is activated by the word-line (WL) decoder.
\endgroup

\begin{figure}[h]
\centering
\centerline{\includegraphics[width=3in]{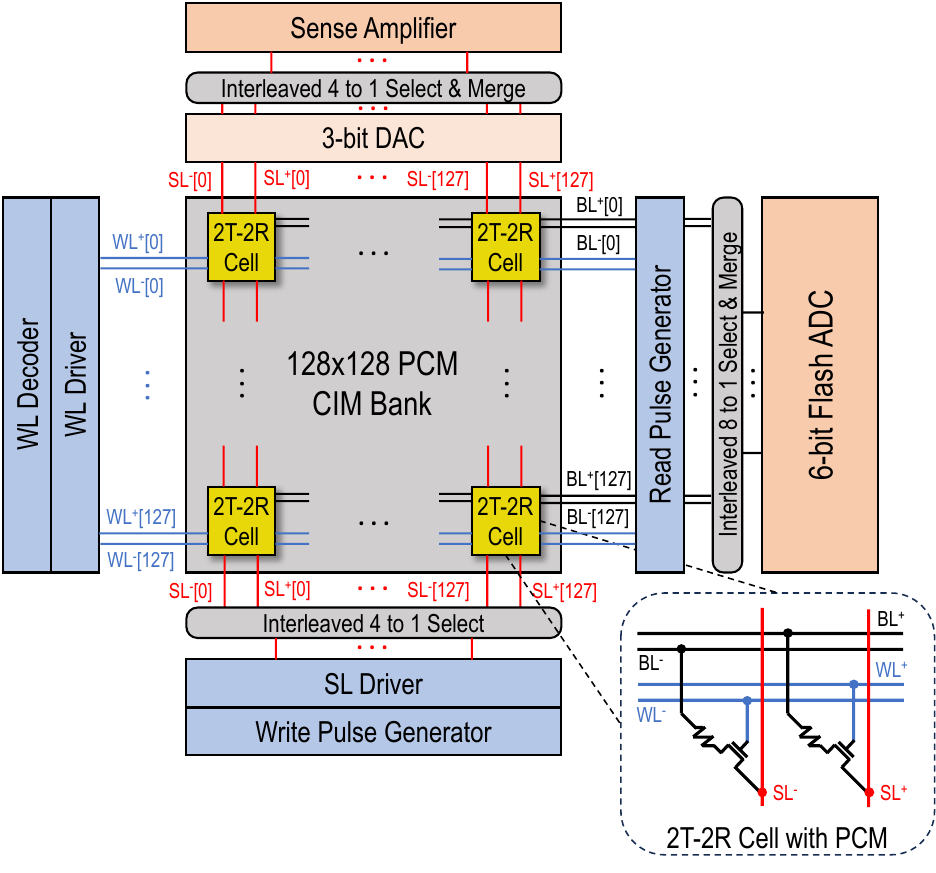}}
\caption{In-memory computing array with 2T2R PCM cell structure.}
\label{fig: cim_array}
\vspace{-0.7cm}
\end{figure}

\begingroup
\noindent \textbf{Normal Read operation:}  To perform a normal read operation, a target row is activated by the WL decoder. 
The pulse generator module creates pulses smaller than 0.4 V to prevent read disturbances in neighboring cells.
The read pulses are applied through BLs of the target row based on the activated WL and the values of a whole row are read at a time through SLs using sense amplifiers.
\endgroup

\begingroup
\noindent \textbf{IMC for clustering:} During clustering, the distances between all HVs need to be calculated. The retrieved HV from the array through the normal read operation serve as an input for the pair-wise distance calculation with all the other HVs in the memory through the IMC operation.
%
During the IMC, the signed input is sent to the array through the SLs using a 3-bit DAC.
To enable parallel computing of the whole array, all the WLs are enabled at the same time to deliver the inputs to multiple rows.
The distances between the input HV and the stored HVs in all the rows are computed at a time via the dot-product operations and generated on the positive and negative BLs (BL+ and BL-) as analog outputs.
These BL voltages are converted to digital values by 6-bit Flash ADCs with 63 dynamic comparators, with each ADC shared across every eight rows.
Each ADC operation takes one cycle to complete, and the entire IMC process requires ten cycles, including input generation overhead via the DACs.

Then, the generated distance matrix is stored in a separate block of PCM memory array. 
This distance matrix is dynamically updated by the near-memory ASIC logic, which manages the merging of data points into a group based on the computed distances. 
The ASIC employs the complete linkage method, where the maximum distance between one element from each of two clusters determines the distance between the clusters. This process iteratively merges the closest clusters and updates the distance matrix accordingly. At each iteration of such operations, the newly generated clusters are written to the memory array through programming operations. Such a programming overhead is mitigated via the device optimization in Section~\ref{sec: pcm devices} and controlling the write-verify iterations.

\noindent \textbf{IMC for DB search:} 
For DB search, an input query HV needs to be compared with all the stored reference HVs simultaneously.
The query HV is converted via the proposed dimension packing, and applied through the SLs as an input of IMC operation.
The PCM array then performs a dot product of the input query vector with all the stored reference HVs simultaneously, same as in the clustering. 
The resulting partial sums from these operations are transmitted to the peripheral ASIC logic. The ASIC then processes these sums and identifies the highest score as the matching candidate, effectively determining the best match between the query and reference HVs.
\endgroup

\subsection{Efficiency vs. Accuracy Trade-offs}\label{sec: trade-off}

We utilize multiple control knobs to manage the efficiency versus accuracy trade-offs, depending on the target application and processing stages.

\noindent \textbf{HD vector dimension:}
The dimension of the HV is an important parameter for controlling the amount of information. A higher HD dimension can be achieved by simply utilizing more memory space, with additional arrays for the distributed storage.

\noindent \textbf{Reconfigurable ADC bits:} 
While the 6-bit ADC is employed with 63 comparators, the effective bit precision can be modulated to be 1 - 6 bits by partially enabling the ADCs to reduce the energy overhead without modifying the hardware.

\noindent \textbf{Write-verify cycles:} 
For each write-verify cycle, the resistance of the cell is read after each write and compared to the target resistance level. An additional pulse is applied  if needed, e.g.,  if the resistance is lower than the target, a  pulse with higher amplitude or iterative pulse is applied, and vice versa. Such a write-verify operation increases the programming overhead by the amount of cycle numbers. Fig.~\ref{fig:write-verify} shows bit error rate vs.  write-verify cycles measured from 100 fabricated devices, averaged over 100 rounds of measurements. The bit error rate decreases as the number of write-verify cycles (and thus write latency) increases.

\vspace{-0.5cm}

\subsection{PCM Device Optimization}\label{sec: pcm devices}
In addition to reconfigurable parameters in Section~\ref{sec: trade-off}, we leverage two different PCM device technologies, based on superlattices of Sb$_2$Te$_3$/Ge$_4$Sb$_6$Te$_7$ and TiTe$_2$/Ge$_4$Sb$_6$Te$_7$, to maximize the efficiency for different use cases. Due to the differences in fundamental material properties, these two technologies provide unique trade-offs between programming energy, retention, and resistance on/off ratio (Supplementary Table~\ref{tab:device_opt}). Sb$_2$Te$_3$/Ge$_4$Sb$_6$Te$_7$ and TiTe$_2$/Ge$_4$Sb$_6$Te$_7$ require programming voltages of 0.65 V to 0.8 V and 0.85 V to 1 V, respectively, with higher voltages needed to program higher resistance levels.
The clustering stage requires iterative programming to update the newly generated clusters periodically, as discussed in Section~\ref{sec: hardware}. 
Given that both types of PCM devices have an endurance of over $10^8$ cycles and clustering typically involves fewer than 100 iterations, resulting in under 100 writes per cell, the system theoretically supports over $10^6$ clustering processes. Furthermore, many iterations involve minor adjustments, particularly in later stages, where the algorithm fine-tunes existing clusters, leading to relatively mild write operations. For this reason, we expect the proposed system to practically support well over $10^6$ clustering processes in realistic scenarios.
The write-intensive nature of clustering makes energy efficiency in programming crucial,
while the retention period of the device can be significantly relaxed. 
 The Sb$_2$Te$_3$/Ge$_4$Sb$_6$Te$_7$ provides higher on-resistance (with a similar on/off resistance ratio of $\approx$100$\times$) and lower programming current, reducing energy consumption in peripheral circuits during the programming operation.
 On the other hand, the DB search requires storing the reference HVs for a long enough retention time to be compared with the input HVs. 
 TiTe$_2$/Ge$_4$Sb$_6$Te$_7$ offers longer retention time and lower error rate at operation temperature (105 $^{\circ}$C) at the cost of 2.6$\times$ higher programming energy. As a result, the MS system for the clustering is implemented using Sb$_2$Te$_3$/Ge$_4$Sb$_6$Te$_7$ PCM while DB search, which requires intensive read operations and longer retention time, is implemented using TiTe$_2$/Ge$_4$Sb$_6$Te$_7$ PCM. %

\begin{figure}
\centering
\centerline{\includegraphics[width=0.36\textwidth]{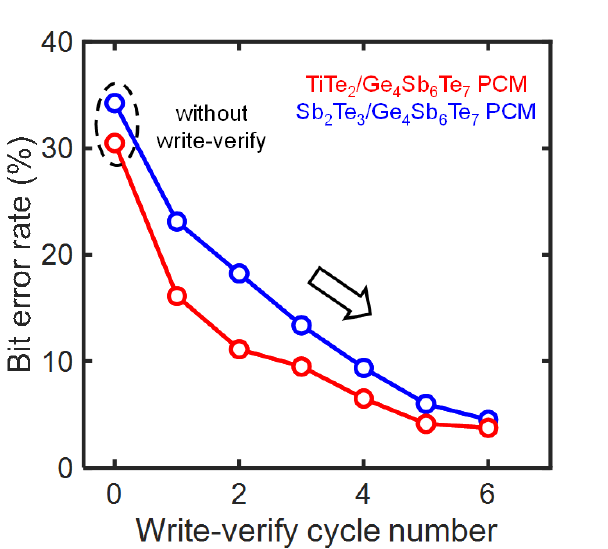}}
\caption{Experimentally measured bit error rate as a function of write-verify cycles for 3-bits per cell.\label{fig:write-verify}}
\vspace{-0.5cm}
\end{figure}

\subsection{Instruction Set} \label{sec: compiler}

Given various parameters discussed in Section~\ref{sec: trade-off}, we developed an Instruction Set Architecture (ISA) to effectively control them from the software. This ISA, detailed in Table~\ref{tab:ISA} manages memory operations in the proposed system %
including read, program, write-verify, and in-memory dot product used in clustering and database search.  The instruction set also configures parameters such as write\_cycles, MLC\_bits, ADC\_bits, and HD\_dimensions given the user's requirements.

\section{Evaluation}\label{sec: evaluation}

This section presents the experimental results of the proposed MS system, highlighting the energy and delay benefits achieved through the proposed optimizations, along with their impact on accuracy for various benchmarks.

\subsection{Experimental Setup}\label{sec:exp_setup}

\textbf{Datasets:}
We evaluate the design using two real-world datasets for clustering and two for DB search, representing  small and large scales. 
For clustering, the following two datasets are employed: 1) a small-scale dataset, PXD001468 \cite{dataset1468} and  2) a large dataset, PXD000561 \cite{dataset0561}.
For DB search, we use: 1) the small-scale iPRG2012 \cite{small_query} and 2) the larger-scale HEK293. The details are described in Supplementary Section~\ref{sec: supp_dataset}. 

\noindent \textbf{Quality Metrics:} For clustering, we evaluate the quality using the cluster spectra ratio, which is the number of clustered spectra divided by the total number of spectra. This metric assesses the clustering capability of each tool while keeping the incorrect clustering ratio fixed. For DB search, we compare the number of total identified peptides given the fixed FDR rate against those identified by other tools.

\begin{table}[t]
\centering
  \caption{SpecPCM Hardware Configurations.}
  \label{tab:config}
  \resizebox{0.49\textwidth}{!}{
  \begin{tabular}{l|l}
    \hline
     \textbf{Component} & \textbf{Configurations}  \\
    \hline
     PCM Array & 128$\times$128 2T2R cells, 3 bits per cell \\
     Flash ADC & 6-bit precision, 16 units each shared between eight rows of cells   \\ 
     DAC & 3-bit precision, 128 units each for one col of cells  \\ 
    SL Gen / Drive &  64 units each shared between four cols of cells  \\
    Read Gen &  Two units for each row of cells, activated for the target row  \\
    WL Decode / Drive  &  8-bit decoder, 256 driver units, two per one row of cells \\
    Sense Amp &  3-bit precision, 32 units each shared between four cols of cells \\\hline
\end{tabular}
}
\vspace{-0.3cm}
\end{table}

\begin{figure}
\centering
\centerline{\includegraphics[width=2.5in]{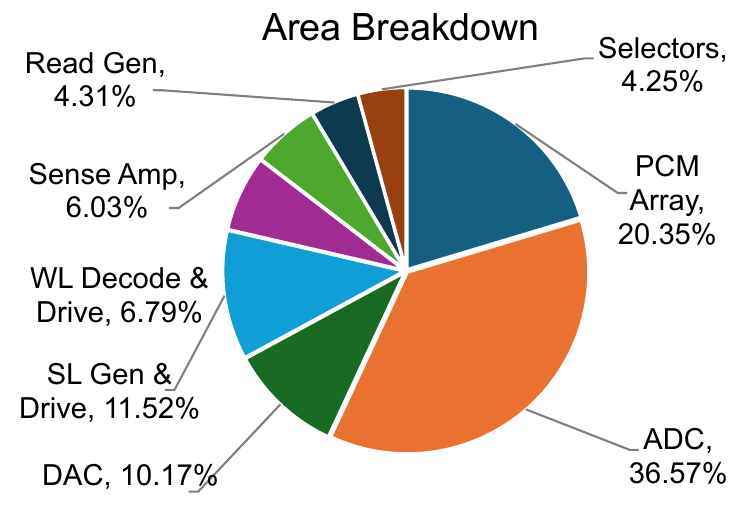}}
\caption{Area breakdown of SpecPCM system.\label{fig:area}}
\vspace{-0.5cm}
\end{figure}

\noindent \textbf{Hardware Configurations:} By default, three write-verify cycles, 3-bit MLC, HD dimension of 8192, and a 6-bit ADC are employed for the DB search. 
On the other hand, the HD dimension is set to 2048 for the clustering.
While the 3-bit MLC and a 6-bit ADC are employed for the clustering similar to the DB search, no write-verify is used for the default setup due to the strong error tolerance of clustering process (see Supplementary Fig.~\ref{trade}(a)). We built an in-house simulator to model our system based on the methodology in Supplementary Section~\ref{sec: sup_experiment setup}. The configuration of each hardware component is detailed in Table~\ref{tab:config}.
The hardware system is built with the CMOS 40~nm process at a target clock frequency of 500~MHz.
The area breakdown is shown in Fig.~\ref{fig:area} indicating high overhead from the ADC. Therefore, an  ADC unit is shared across eight rows of cells to minimize the area overhead. 
The detailed power breakdown is shown in Supplementary Table~\ref{tab:comp_energy}.
%

\noindent \textbf{Baseline designs:}
For clustering, we evaluate our approach against four state-of-the-art tools, including Falcon \cite{bittremieux2021falcon}, msCRUSH \cite{wang2018mscrush}, HyperSpec \cite{xu2023hyperspec}, and SpecHD \cite{pinge2024spechd}.  For DB search, we compare our system against ANN-SoLo \cite{ANN_SoLo'23} on GPU and HyperOMS \cite{kang23hyperoms} on GPU, and IMC accelerators including RRAM-based\cite{fan2024efficient} and 3D NAND-based\cite{hsu2023storage}. All DB search results are evaluated at a fixed 1\% FDR threshold, the same as existing works. The baseline systems are tested on NVIDIA GeForce RTX 4090 GPU with 24 GB VRAM and Intel Core i7-11700K CPU with 64 GB of RAM. 

\subsection{Experiment Results}


\textbf{Search Quality:}
Fig.~\ref{clustering_acc} shows that the SpecPCM outperforms existing tools such as Falcon \cite{bittremieux2021falcon} and msCRUSH \cite{wang2018mscrush}, while delivering performance comparable to HyperSpec \cite{xu2023hyperspec} across all configurations, including single-level-cell (SLC), 2-bit MLC (MLC2), and 3-bit MLC (MLC3). With an incorrect clustering ratio of up to 1.5\%, SpecPCM achieves around 60\% clustered spectra ratio. 
Compared to SLC, the performance reduction in both MLC2 and MLC3 is minimal, indicating that the dimension-packing method has a negligible impact on accuracy.

\begin{figure}[tp]
\centerline{\includegraphics[width=3.1in]{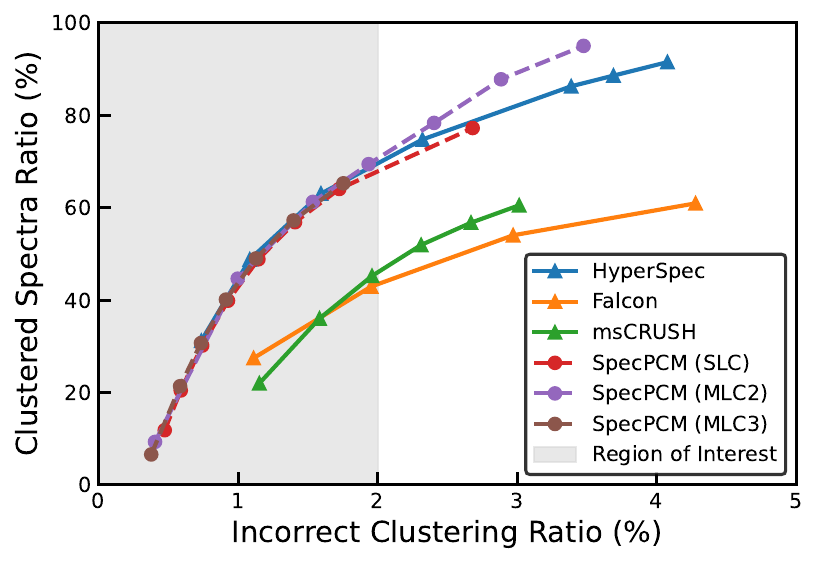}}
\caption{Clustering quality for  PXD000561 dataset. Given incorrect clustering ratio, higher clustered spectra ratios indicate superior clustering performance. The region of interest is  an incorrect clustering ratio of less than 2\%.}
\label{clustering_acc}
\vspace{-0.5cm}
\end{figure}

In DB search, we validate the functionality of SpecPCM by comparing it against existing tools.
Supplementary Fig.~\ref{oms_acc} illustrates the total number of peptides identified using the large-scale HEK293 dataset, and  Fig.~\ref{sup_fig: hek_venn} visualizes the identified data points for the specific query b1931 in a Venn diagram.
SpecPCM demonstrates higher search quality compared to SpectraST \cite{ma2014spectraST} and comparable performance to HyperOMS \cite{kang23hyperoms}.  
Although ANN-SoLo identifies the highest number of peptides, it comes at the cost of significantly higher power consumption and latency (Table~\ref{table:DB_latency_table}).
While the proposed SpecPCM provides balanced accuracy, parameters such as HD dimensions and others can be further adjusted to enhance search quality at the cost of increased energy and latency (see Supplementary Fig.~\ref{trade}a and \ref{trade}b). 

\begin{figure}[tp]
\hspace{0.3cm}
{\includegraphics[width=0.42\textwidth]{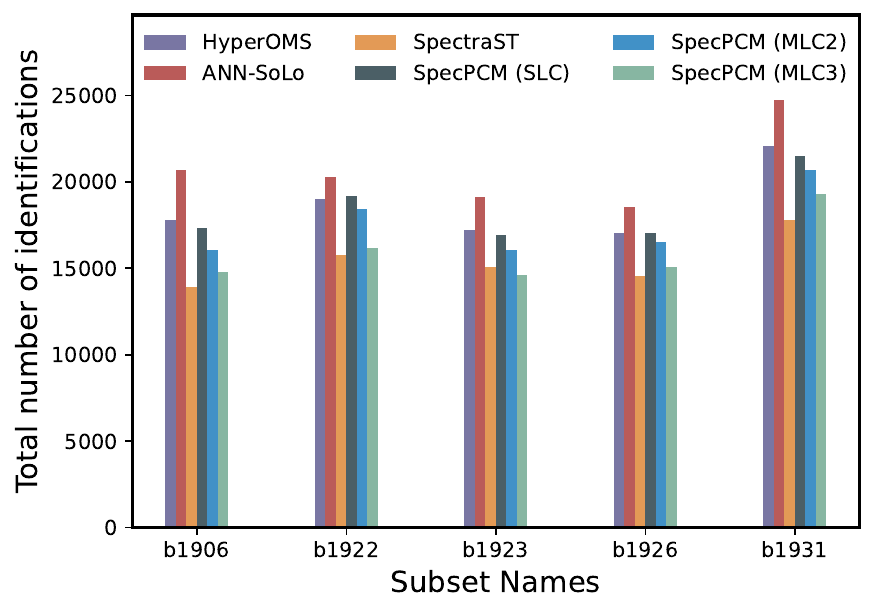}}
\caption{DB search quality for the HEK293 dataset, showing the number of identified peptides for each subset.}
\label{oms_acc}
\vspace{-0.5cm}
\end{figure}

\noindent \textbf{Speedup and Energy Efficiency:}
SpecPCM achieves up to 82$\times$ speedup in clustering and up to 143$\times$ speedup in DB search compared to traditional CPU and GPU tools on large datasets as shown in Table~\ref{table:clus_latency_table} and Table~\ref{table:DB_latency_table}, respectively. This performance gain is attributed to the reduced data movement in SpecPCM, which is the dominant overhead in traditional CPU/GPU architectures.
Additionally, SpecPCM fully leverages MLC technology due to the dimension packing.
This packing enables SpecPCM to leverage the benefits of 3-bit MLC,
boosting computation throughput by 3$\times$ compared to the SLC implementation. The energy consumption of SpecPCM is 3.27 J on the PXD00561 dataset for the entire clustering process and 0.149 J for the DB search to process a subset of data, which requires 46,665  query processing on average for the HEK293 dataset. Given that GPU-based tools typically operate at an average power of 450 W with longer delays, SpecPCM is expected to provide an energy efficiency improvement of four orders of magnitude.

When compared to prior MS systems based on 3D NAND~\cite{hsu2023storage} and RRAM~\cite{fan2024efficient}, SpecPCM achieves a speedup of 2.96$\times$ to 24.9$\times$ on the DB search task as shown in Table~\ref{table:DB_latency_table}. While RRAM shows strong potential, it suffers from high voltage requirements for programming and increased noise, leading to larger peripheral circuits and a limited number of activated rows in the array at a time. In contrast, NAND flash has slower read latency compared to PCM due to its serial array structure.

\begin{table}
  \caption{Clustering Speedup vs. Prior Works.}
  \centering
  \label{table:clus_latency_table}
  \resizebox{0.49\textwidth}{!}{
  \begin{tabular}{c|ccccc}
    \hline
    Tools & Falcon & msCRUSH & HyperSpec &SpecHD & SpecPCM \\ \hline
    Hardware & CPU & CPU & GPU &FPGA  & TSMC 40nm \\ \hline
    Dataset & \multicolumn{5}{c}{PXD001468} \\ \hline
    Latency & 573s & 358s & 38s &13.17s& 5.46s  \\
    Speedup & 1$\times$ & 1.6$\times$ & 15.1$\times$&43.5$\times$& 104.94$\times$  \\
    \hline
    Dataset & \multicolumn{5}{c}{PXD000561} \\ \hline
    Latency & 134 min & 42 min & 17 min &179s& 98.4s \\
    Speedup & 1$\times$ & 3.2$\times$ & 7.9$\times$ &44.9$\times$ &81.7$\times$ \\ \hline
    \end{tabular}
    }
    \vspace{-0.3cm}
\end{table}

\begin{table}
  \caption{DB Search Speedup vs. Prior Works.}
  \centering
  \label{table:DB_latency_table}
  \resizebox{0.49\textwidth}{!}{
  \begin{tabular}{c|ccccc}
    \hline
    Tools & ANN-SoLo & HyperOMS & RRAM & 3D NAND & SpecPCM \\ \hline
    Hardware & CPU-GPU & GPU & 130nm & ASAP 7nm & TSMC 40nm \\ \hline
    Dataset & \multicolumn{5}{c}{iPRG2012} \\ \hline
    Latency & 6.45s  & 2.08s  &  1.22s  & 0.145s  & 0.049s  \\
    Speedup & 1$\times$ & 3.1$\times$ & 5.3$\times$ & 44.2$\times$ & 131.63$\times$ \\
    \hline
    Dataset & \multicolumn{5}{c}{HEK293} \\ \hline
    Latency & 45.14s & 10.4s & - & -  & 0.316s \\
    Speedup & 1$\times$ & 4.34$\times$ & - & - & 142.84$\times$ \\
    \hline
  \end{tabular}}
  \vspace{-0.3cm}
\end{table}

\noindent \textbf{Accuracy and efficiency trade-offs:} 
Our PCM-based IMC accelerator allows the ISA-based control of following parameters that impact the performance.
In this section, we explore various combinations of these parameters to analyze their impacts on the quality of MS analysis. 

\noindent (1) Bits per cell: Increasing the number of bits per cell enhances memory and compute density. However, this leads to higher error rates, degrading the MS performance as illustrated in Fig.~\ref{clustering_acc} and Fig.~\ref{oms_acc}. In clustering, the accuracy decreases from 60.57\% with SLC to 59.80\% and 59.54\% with 2-bit and 3-bit MLC, respectively, at an incorrect clustering ratio of 1.5\%. In contrast, the DB search shows a more pronounced drop in quality, highlighting its greater sensitivity to noise.

\noindent (2) HD dimension: Higher HD dimensions generally enhance performance at the cost of linearly increased storage and processing delay and energy  (Supplementary Fig. \ref{sup_fig: clus_D}, ~\ref{sup_fig: oms_D}).

\noindent (3) Write-verify cycles: DB search operations are predominantly read-oriented, allowing the cost of write-verify operations on reference HVs to be effectively amortized. This means that a higher number of write-verify cycles can be used to ensure accurate storage. In contrast, spectral clustering involves frequent write operations to update the distance matrix. Thus, increasing the number of write-verify linearly increases the latency and energy consumption. 
Fortunately, the performance in terms of clustered spectra ratio remains largely unaffected by the number of write-verify cycles, as shown in Supplementary Fig. \ref{trade}(a). As a result, no write-verify cycles are used for clustering in this work.

\noindent (4) ADC resolution: Since HD vectors have positive and negative values with an equal probability, partial sums are likely to approach near-zero values. 
In addition, HD computing has a high inherent error resiliency.
For this reason, the performance gracefully degrades as ADC bit precision reduces. 
Therefore, a 4-bit flash ADC can achieve roughly 4$\times$ less area and energy compared to a 6-bit flash ADC at the marginally degraded accuracy (Supplementary Fig.~\ref{trade}(b)).

By tuning the above parameters using the provided instruction sets, we can optimize the performance and efficiency of the proposed SpecPCM, highlighting its flexibility and adaptability given the target specification.

\section{Conclusion}
This work presents a full-stack effort for a PCM-based analog in-memory computing (IMC) platform that significantly enhances energy and delay efficiency for mass spectrometry (MS) analysis, addressing both clustering and DB search. In consideration of clustering, which requires frequent memory updates, we utilize superlattice-material-based PCM for low-voltage programming and employ different PCM materials to optimize the distinct requirements of clustering and DB search.
At the algorithm level, we introduce dimension packing to compress binary data into multi-level cells, maximizing storage and IMC processing efficiency. Additionally, we leverage a hyperdimensional computing (HD) platform for its strong error tolerance and parallel processing capabilities with simple operations. The proposed instruction set architecture (ISA) enables software-driven trade-offs between accuracy and efficiency by controlling various hardware and algorithmic parameters.
Experimental results based on a 40 nm CMOS process demonstrate that SpecPCM achieves speedups of up to 82$\times$ for clustering and 143$\times$ for DB search, alongside a four-orders-of-magnitude improvement in energy efficiency while maintaining accuracy. These full-stack efforts underscore the potential for greater efficiency through automated compiler-based optimizations and pave the way for future work in many data-intensive tasks, overcoming the non-idealities of various emerging technologies.

%


\bibliographystyle{ieeetr}
\bibliography{references.bib}

\newpage
\clearpage
\section{Supplementary Material}

\setcounter{subsection}{0}  
\renewcommand{\thesubsection}{S.\Alph{subsection}}
\renewcommand{\thesubsectiondis}{S.\Alph{subsection}}

\setcounter{figure}{0}  
\renewcommand{\thefigure}{S\arabic{figure}}

\subsection{Dataset} \label{sec: supp_dataset}
All spectra are pre-processed following the existing methodology in \cite{kang23hyperoms, xu2023hyperspec}. 
All MS data, spectral libraries, pre-processed spectra, and identification results are available for download from the MassIVE repository under the dataset identifier MSV000091183 \cite{MassIVE}.

For clustering, the following two datasets are employed: 1) a small-scale dataset, PXD001468 \cite{dataset1468}, which contains 1.1 million spectra from kidney cell samples with a total size of 5.6 GB, and 2) a large dataset, PXD000561 \cite{dataset0561}, which contains a draft map of the human proteome with 21.1 million spectra totaling 131 GB.
For DB search, we use: 1) the small-scale iPRG2012 \cite{small_query} as the query (total spectra: 15,867) and the human HCD yeast library \cite{small_ref} as the reference database (total spectra: 1,162,392), and 2) the larger-scale HEK293 (Human Embryonic Kidney 293) \cite{large_query}, which includes multiple different subsets b1906$\sim$ 1931 (total spectra per subset: 46,665 on average), as the query and the human spectral library \cite{large_ref} (total spectra: 2,992,672) as the reference library.
\vspace{-0.3cm}

\subsection{Experiment setup}
\label{sec: sup_experiment setup}
\noindent \textbf{Hardware Configurations:} 
For the ASIC blocks, we initially implement the logic in C++ and subsequently generate the RTL code using high-level synthesis (HLS). The RTL code is then synthesized using Cadence Genus with the CMOS 40~nm process PDK to meet the target clock frequency of 500~MHz.
The ASIC area for the encoder is 44 ${\mu}m^2$  and 69 $\mu m^2$ for other components, both of which are negligible (less than 0.5\%) compared to that of the memory arrays.
Each 2T2R cell has an area of 0.5${\mu}m^2$. We use the power and area data of DAC in \cite{saberi2011ADC}. For all the other blocks, the schematic and layout are designed using 40~nm CMOS technology in Cadence Virtuoso, and their corresponding power and area are measured from the post-layout simulations. 
The component-level energy and area  are summarized in Table~\ref{tab:comp_energy}.
Most operations of the components listed in Table~\ref{tab:comp_energy} complete within one cycle, whereas the programming of a PCM array takes 20 ns (10 cycles).

\noindent \textbf{PCM device:} PCM devices in this work have TiN bottom electrodes with 40 nm diameter (Fig.~\ref{sup_fig: device_image}) \cite{wu2024pcm}. Phase change superlattices are deposited using magnetron sputtering (detailed fabrication process can be found in \cite{wu2024pcm}). Device noise and resistance drift are measured based on protocols described in \cite{wu2023drift}. Resistance drift coefficient is extracted from resistance vs. time measurement using power-law model. 

\noindent \textbf{Noise model:} 
We fit PCM resistance measurements to a normal distribution to derive the standard deviation ($\sigma$). This normal distribution with $\sigma$ models the noise effect on the weights stored in the memory. In the simulation, noise adjustment is applied to validate the impact of the noise on the performance, e.g., $\hat{W} = W \times (1 + \eta)$, where $\eta \sim \mathcal{N}(0, \sigma^2)$. Here, \( W \) represents the error-free stored value, and \( \hat{W} \) denotes the erroneous read value, with \( \mathcal{N} \) is a normal distribution.

\begin{figure}[ht]
    \centerline{\includegraphics[width=0.39\textwidth]{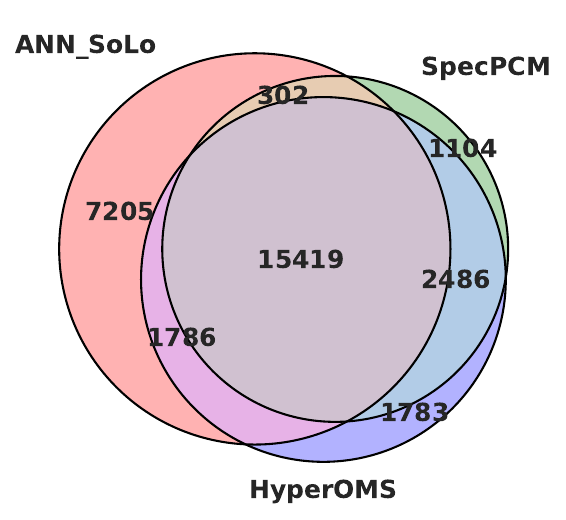}}
    \caption{DB search result: an example of identified peptides in Venn diagram with MLC3 for the dataset HEK293 b1931. The majority of peptides detected by SpecPCM can also be found by other tools, indicating reliability of the SpecPCM results.}
    \label{sup_fig: hek_venn}
\end{figure}

\begin{figure}[!ht]
    \centerline{\includegraphics[width=0.39\textwidth]{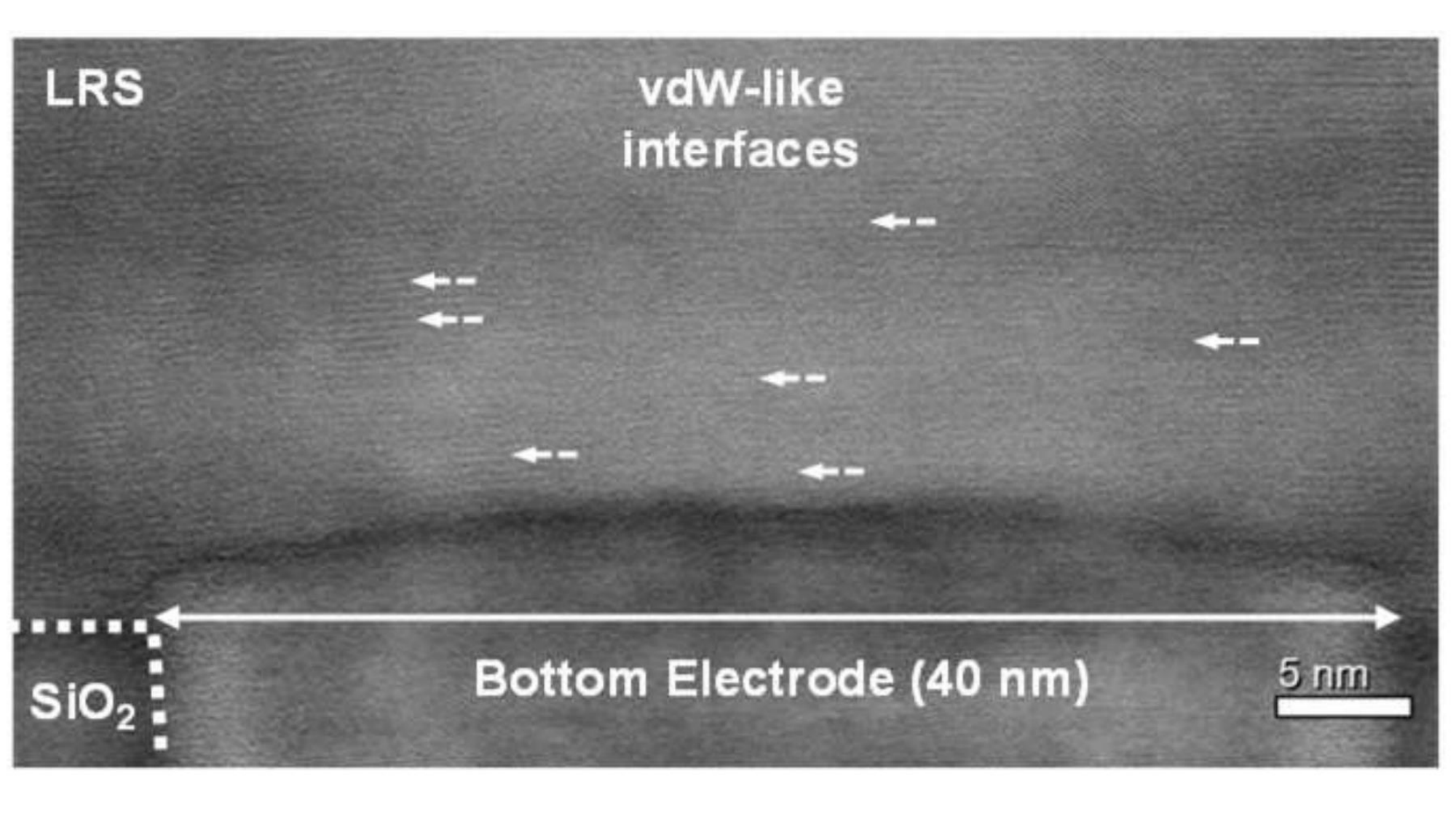}}
    \caption{Transmission Electron Microscopy (TEM) image of a TiTe$_2$/Ge$_4$Sb$_6$Te$_7$ device used in this work, showing 40 nm bottom electrode and superlattice-like phase change materials.}
    \label{sup_fig: device_image}
\end{figure}

\setcounter{table}{0} 
\renewcommand{\thetable}{S\arabic{table}}

\begin{table}[!ht]
\centering
  \caption{Measured parameters of PCM device technology.}
  \label{tab:device_opt}
  \resizebox{0.45\textwidth}{!}{
  \begin{tabular}{l|c|c}
    \hline
     \textbf{Technology} & \textbf{Sb$_2$Te$_3$ / Ge$_4$Sb$_6$Te$_7$} & \textbf{TiTe$_2$ / Ge$_4$Sb$_6$Te$_7$} \\
    \hline
     Programming current ($\mu$A) & 80 & 160  \\
     Programming voltage (V)  & 0.7 & 0.9  \\
     Programming energy (pJ) & 1.12 & 2.88  \\ 
     Retention at 105$^\circ$C (hour)  &  30 & $>10^5$   \\ 
     Low resistance state (kOhm) &  30 & 10  \\
     Resistance on/off ratio  &  150 & 100 \\\hline
\end{tabular}
}
\end{table}

\begin{figure}[!ht]
	\centering
	\begin{subfigure}[t]{0.49\textwidth}
		\centering
		\includegraphics[width=0.8\textwidth]{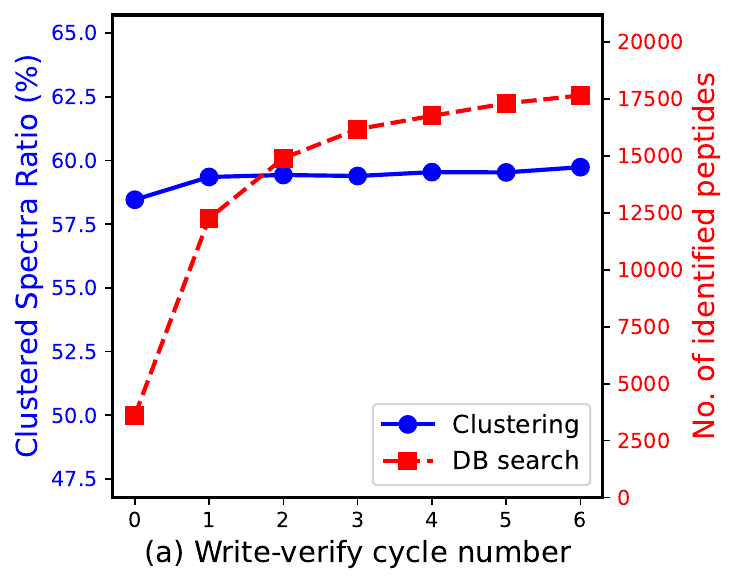}
  \vspace{-8pt}
        \label{trade_write}
	\end{subfigure}
    \hfill
	\begin{subfigure}[t]{0.49\textwidth}
		\centering
		\includegraphics[width=0.78\textwidth]{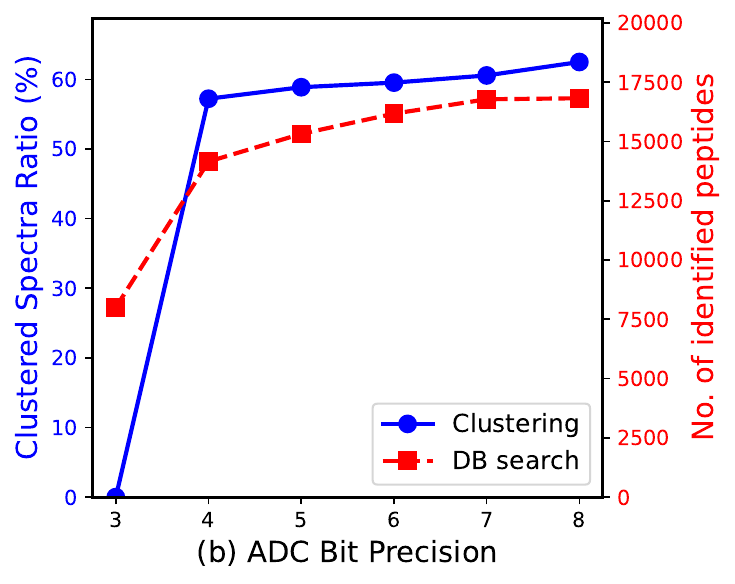}
        \label{trade_adc}
	\end{subfigure}
    \caption{Accuracy and efficiency trade-offs. (a) quality vs. write-verify cycle number and (b) quality vs. ADC bit precisions.}
    \label{trade}
\end{figure}

\begin{table*}[t]
\centering
  \caption{Instruction Set Architecture (ISA) for IMC control.}
  \label{tab:ISA}
  \resizebox{1.0\textwidth}{!}{
  \begin{tabular}{c|c}
    \hline
     \textbf{Instructions} & \textbf{Description}  \\
    \hline
     STORE\_HV (data, arr\_idx, col\_addr, row\_addr, {MLC\_bits}, write\_cycles) & \makecell{ PCM$[$arr\_idx, col\_addr, row\_addr$]$ $\leftarrow$ data.\\ \textbf{write\_cycles} defines the number of write verify cycles\\  \textbf{MLC\_bits} defines the number of bits used by dimension packing for MLC} \\
    \hline
     READ\_HV (data\_size, arr\_idx, col\_addr, row\_addr, {MLC\_bits}) &\makecell{ buffer$\leftarrow$ PCM$[$arr\_idx, col\_addr, row\_addr$]$ \\\textbf{MLC\_bits} is configured same as above }\\ 
    \hline
     MVM\_COMPUTE (row\_addr, num\_activated\_row, {ADC\_bits}, {MLC\_bits})& \makecell{ Compute distance through Matrix-Vector Multiplication (MVM) at PCM$[$row\_addr$]$ \\ \textbf{num\_activated\_row} defines size of activated weight matrix\\ \textbf{ADC\_bits} defines the resolution of the Flash ADC }\\ \hline
\end{tabular}}
\end{table*}

\begin{figure}[!ht]
    \centerline{\includegraphics[width=0.3\textwidth]{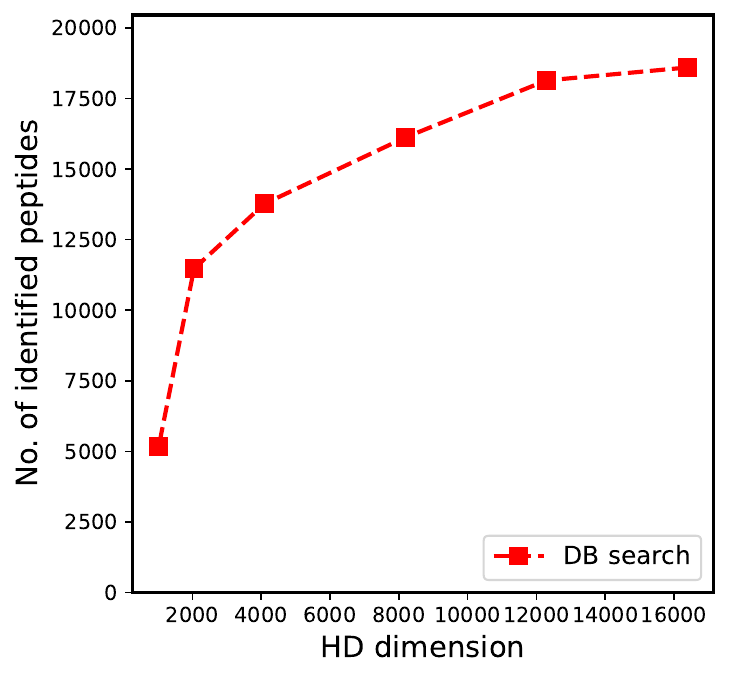}}
    \caption{DB search quality vs. HD dimension.}
    \label{sup_fig: oms_D}
\end{figure}

\begin{table*}[!ht]
\centering
  \caption{Power and area of components at 40~nm CMOS process.}
  \label{tab:comp_energy}
  \begin{tabular}{l|c|c|c|c}
    \hline
     \textbf{Component} & \textbf{Unit Power (${\mu}W$)} & \textbf{Unit Area (${\mu}mm^2$)} & \textbf{Total Power ($mW$)} & \textbf{Total Area ($mm^2$)} \\
    \hline
     PCM Array & 0.22 & 0.5 & 3.58 & 0.0082  \\
     Flash ADC & 320 & 920 & 5.12 & 0.0147  \\ 
     DAC & 6.56 & 32 & 0.84 & 0.0041  \\ 
    SL Gen / Drive & 52.5 & 72.47 & 3.36 & 0.0046  \\
    Read Gen & - & - & 0.51 & 0.0018  \\
    WL Decode / Drive  & 4.05 & 10.68 & 1.04 & 0.0027 \\
    Sense Amp & 20 & 75.9 & 0.64 & 0.0024 \\
    Selectors & - & - & 0.50 & 0.0017  \\\hline
    \textbf{Total} & \textbf{-} & \textbf{-} & \textbf{15.59} & \textbf{0.0402} \\\hline
\end{tabular}
\end{table*}

\begin{figure}[!ht]
    \centerline{\includegraphics[width=0.3\textwidth]{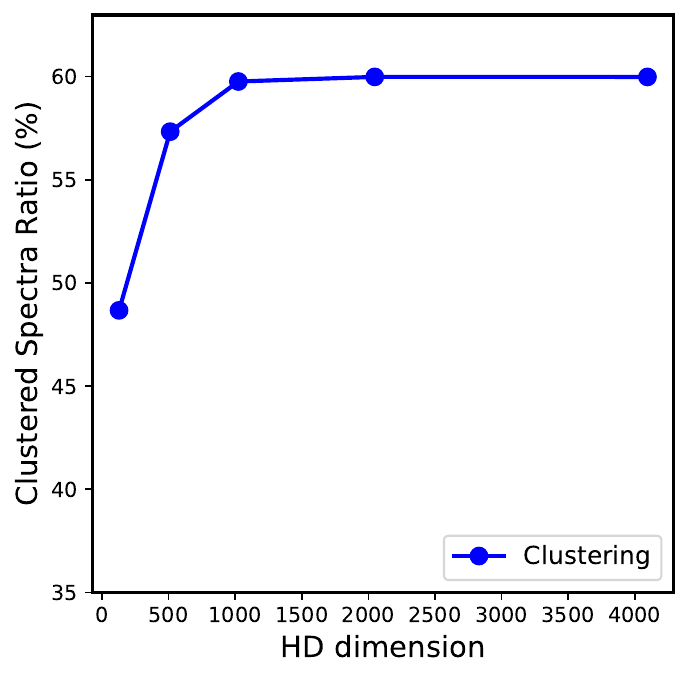}}
    \caption{Clustering quality vs. HD dimension.}
    \label{sup_fig: clus_D}
\end{figure}

\newpage
\clearpage
\onecolumn

\end{document}